\documentclass{elsart}
\usepackage{graphicx}
\usepackage{amssymb}

\renewcommand{\vec}[1]{\mathbf{#1}}
\newcommand{\jpg}{J.\ Phys.\ G: Nucl.\ Part.\ Phys. }

\begin{document}
\hspace*{\fill}%
\begin{minipage}{7cm}
\flushright
UMIST/Phys/TP/98-7
\end{minipage}

\begin{frontmatter}
\title{Translationally invariant treatment of pair correlations in nuclei: II.
Tensor correlations}

\author[UMIST]{R. F. Bishop}, \author[Fisato]{R. Guardiola},
\author[Fisato]{I. Moliner}, \author[IFIC]{J. Navarro}
\author[IFIC]{M. Portesi}, \author[Balears]{A. Puente}, 
and \author[UMIST]{N. R. Walet\thanksref{emailNRW}}

\address[UMIST]{Department of Physics, UMIST, P.O. Box 88,
Manchester M60 1QD, U.K.}

\address[Fisato]{Dpto. de F\'{\i}sica At\'omica, Molecular y Nuclear,
Universitat de Valencia, Avda. Dr. Moliner 50, E-46100 Burjassot,
Spain}

\address[IFIC]{IFIC (Centre Mixt CSIC -- Universitat de Valencia),
Avda. Dr. Moliner 50, E-46100 Burjassot, Spain}

\address[Balears]{Departament de F\'{\i}sica, Universitat de les Illes Balears,
E-07071 Palma de Mallorca, Spain}

\thanks[emailNRW]{Email: Niels.Walet@umist.ac.uk}

\begin{abstract}
We study the extension of our translationally invariant treatment of
few-body nuclear systems to include tensor forces and correlations.
It is shown that a direct application of our method is not as
successful for realistic V6 interactions as our previous results for
V4 potentials suggested. We investigate the cause in detail for the
case of $^4$He, and show that a combination of our method with that of
Jastrow-correlated wave functions seems to be a lot more powerful, thereby
suggesting that for mildly to strongly repulsive forces such a hybrid
procedure may be an appropriate description.
\end{abstract}
\end{frontmatter}


\section{Introduction}

The last few years have seen an impressive progress in the microscopic
description of light, $A\lesssim 16$, nuclei directly in terms of
bare realistic nucleon-nucleon interactions.  Among the most important
developments are the variational and quantum Monte Carlo methods
\cite{LANL1,LANL2,LANL3,PWP}, the stochastic variational approach
\cite{Kalman}, the fully-converged calculations within the Faddeev and
Faddeev-Yakubowsky theories \cite{Bochum}, and analogous calculations
in the correlated hyperspherical harmonics expansions \cite{Pisa}.

Some of these approaches are limited by their very nature to three- or
four-body systems. On the other hand, the quantum Monte Carlo method,
the stochastic variational method and the correlated hyperspherical
harmonics expansions are being extended to nuclear systems with up to
around 8 or 10 particles.  At present the main limitations are both
computer speed and computer memory, and even dealing with one
additional particle requires an impressive growth of computing power.
For example, Table 1 in Ref.\ \cite{Pieper} gives an indication of the
computational requirements of quantum Monte Carlo calculations.

At the same time
we have been trying to apply what is essentially a many-body
technique (the coupled-cluster method or CCM) to light systems.
When dealing with finite systems of mutually interacting particles, 
it is important to separate the
centre-of-mass (CM) motion exactly from the intrinsic motion.
In the framework of the CCM it has been shown that it is possible
to do so by using very special correlation operators \cite{Bis90}.
Inspired by the CCM we
have constructed a variational method that is explicitly translationally
invariant, but which includes 
only a linearised correlation operator. 
This procedure, the translationally invariant
configuration-interaction (TICI) method \cite{Bis90,Bis91}, 
is the subject of this paper.
 The procedure  has mainly been used in a two-body implementation,
called  TICI2.
In recent investigations \cite{Gua96,CMT21} this method has been extended to
deal with saturated nuclei within the $0p$ shell, using state-dependent
(SD) pair correlations between nucleons. Calculations have been
performed  for various V4 nucleon-nucleon interactions.

Another way to appreciate the TICI2 method is to consider it
as a special formulation
of the configuration-interaction (CI) method. The main difference with
conventional CI calculations lies in the way the interacting configurations
are selected. In the TICI2 approach these are restricted to the most
general single-pair excitations which leave the CM coordinate
unaffected, and are {\em not} limited to a few major shells, as in
conventional shell model approaches.
 These configurations are very similar to those
considered in the so-called {\em potential basis} in the
hyperspherical harmonics formalism \cite{hh}.

For systems of bosons it has been possible to extend
the TICI2 method to deal with the full 
second-order CCM problem \cite{cc2}.
In this approach, called TICC2, the ground-state wave function
contains the excitation of 
 all possible
{\em independent} pairs. In the specific case of $^4{\rm He}$ 
with  Wigner forces it was found that there
is no significant improvement when going from TICI2 to TICC2,
indicating the lack of importance of two-pair
excitations by contrast with triplet excitations. 
One should beware, however, that there is no guarantee that this statement
holds for realistic nuclear forces with their complicated operatorial
structure.

In our previous work \cite{Gua96} we have shown how well the TICI2 method can
deal with  V4 forces and V4 correlations, defined explicitly in 
Sec.~\ref{sec:formalism}. In this paper we shall
discuss  the extension to tensor forces for  $0p$
shell nuclei.  The TICI2 formalism  is presented in
Sec.~\ref{sec:formalism}, together with a hybrid method that adds scalar
Jastrow correlations to it. 
In Sec.~\ref{sec:results} we present our results, with a special
emphasis on the case 
 of $^4$He nucleus. 
 We find that in order to get a good description
of the $^4$He nucleus we need to mix the operatorial structure of the
TICI2 wave function with a scalar Jastrow-like factor.  We show that
this improves  V4 calculations
as well, but not as dramatically
as in the V6 case.
In Sec.~\ref{sec:conclusions} we draw some conclusions about the new
hybrid method.

\section{Formalism}\label{sec:formalism}
\subsection{Linear pair correlations}
In the SUB2 truncation of the CCM, the ground-state wave function
of an $A$-particle system is parametrised as
\begin{equation}
\vert\Psi_0\rangle= \exp (S_1+S_2) \, \vert\Phi\rangle ,
\end{equation}
where $\vert\Phi\rangle$ is an independent-particle wave function with
appropriate quantum numbers, and the cluster operators $S_1$ and $S_2$
promote one and two particles, respectively, out of the orbits occupied in the
reference state $\vert\Phi\rangle$ to otherwise empty ones.
If the (uncorrelated) reference state is taken to be a
Slater determinant of harmonic oscillator (HO) single-particle
wave functions, the CM motion can be factorised exactly. The remainder
of the wave function is thus translationally invariant. If
one wishes to preserve this translational invariance in the state
$\vert\Psi_0\rangle$,
the excitations produced by the operators $S_1$ and $S_2$ can not be
chosen independently. As has been demonstrated in Ref. \cite{Bis90},
one has to construct a combination of the two operators $S_1$
and $S_2$, denoted by
$S^{(1,2)}$, which mixes both one- and two-body correlations in such
a way that translational invariance is preserved.


In the TICI2 method one expands the exponential
of $S^{(1,2)}$, and keeps only the linear 
term. This can be 
written most concisely in  the coordinate
representation. Here  the TICI2 wave function takes on the very appealing
form
\begin{equation}
\Psi_0({\bf r}_1,\ldots,{\bf r}_A) = 
\sum_{i<j=1}^A f(r_{ij}) \; \Phi({\bf r}_1,\ldots,{\bf r}_A).
\end{equation}
This form is  very suitable  for interactions
that are given in coordinate space, such as those obtained from
$NN$ phase shifts. In order to deal with realistic forces we have
to allow for state
dependence of the pair correlation function $f$. This
is done by promoting $f$ to be an operator, and expanding it in
an appropriate operatorial basis, $f(r_{ij}) = \sum_p f^p (r_{ij})
\Theta^p_{ij}$. The basis of operators to be used is generally chosen
to be consistent with that of the internucleon potential employed. For
interactions with V4 structure (i.e, central, but spin-isospin dependent),
the operators used are $\{\Theta^p_{ij}\} =
\{1,P^{\sigma},P^{\tau},P^{\sigma}P^{\tau}\}$, where $P^{\sigma}$ and
$P^{\tau}$ are the usual spin- and isospin-exchange operators,
respectively. When taking into account
V6 interactions, we need to use  a basis
which also includes both the tensor operators 
$\mathcal{S}_{ij}$ and $\mathcal{S}_{ij} P^\tau$.

The uncorrelated reference state is a single Slater
determinant for the case of spin-isospin saturated systems.
This can be uniquely specified 
by the occupied harmonic oscillator states.
Rather than employing the standard $nls$ basis,
it will be convenient to work in a Cartesian basis where the states
are labelled by three harmonic oscillator quantum numbers 
($n_x,n_y,n_z$). The closed shell nuclei $^4$He and $^{16}$O are
described by  fully filled $n\leq 0$ and $n\leq1$ states; it also
allows for the use of non-spherical reference 
states for the $^8$Be and $^{12}$C nuclei, 
$\vert (0,0,0)^4(0,0,1)^4\rangle$ and 
$\vert (0,0,0)^4(1,0,0)^4(0,1,0)^4\rangle$, respectively, which 
has proven to be better than considering a spherical reference
state \cite{Gua96}.

In the specific case of the $^4$He nucleus,
the reference wave function $|\Phi\rangle$ 
is fully space-symmetric and so the action of the spin-exchange 
operator is equal to the isospin-exchange one, with opposite sign 
(i.e., space symmetry makes $P^\sigma_{ij}P^\tau_{ij}\equiv -1$). Thus in 
this case the basis for the correlations is 
overcomplete and to avoid problems when solving for the correlation
functions $f^p$ we discard the redundant terms, and work in a basis of
only three operators, $1$, $P^\sigma$ and $\mathcal{S}$.

\begin{figure}
\begin{center}
\includegraphics[width=0.7\textwidth]{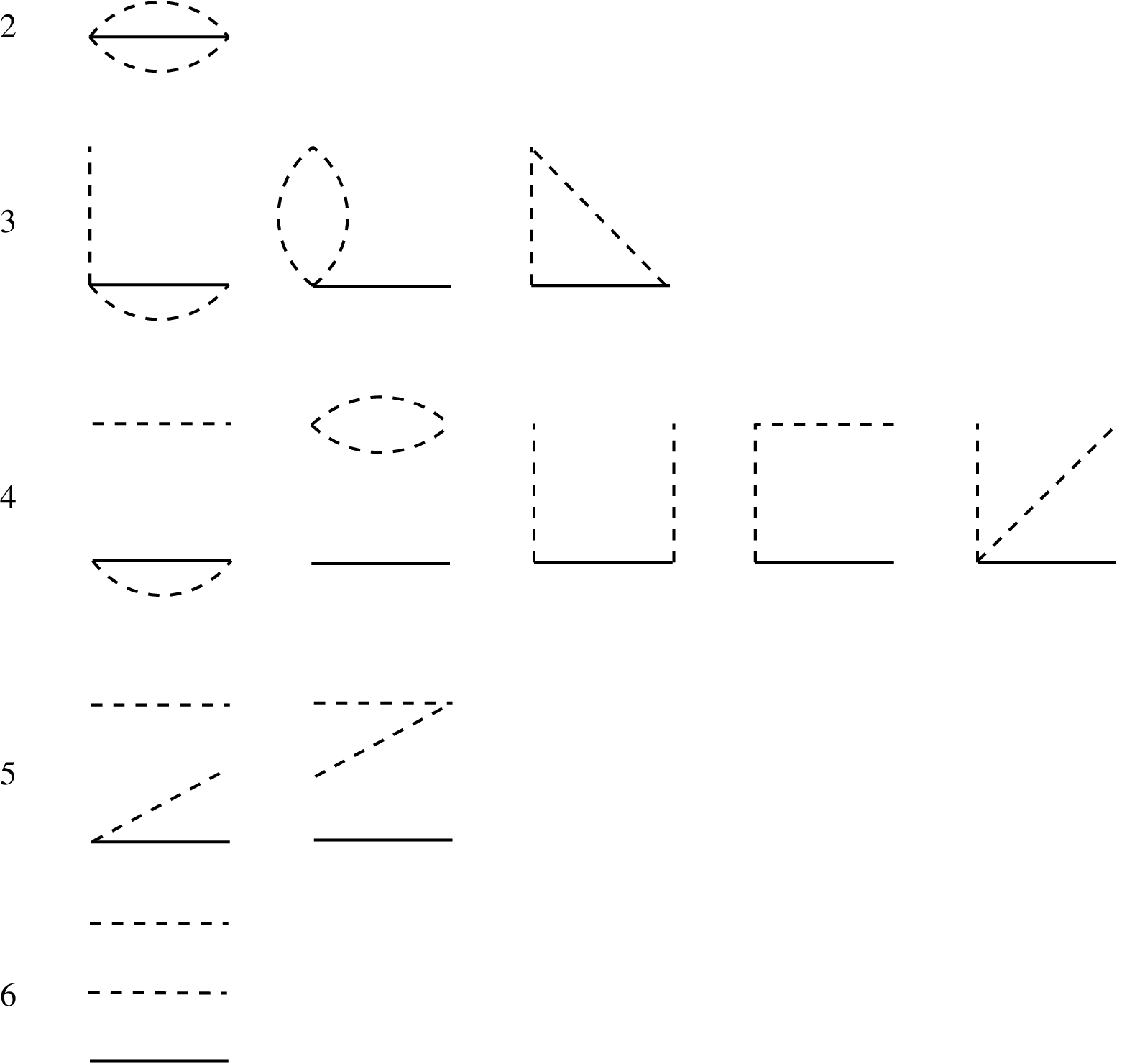}
\end{center}
\caption{A graphical representation of the diagrams contributing to
the potential energy, ordered according to, and labelled by, the
number of particles $N$ involved.  A solid line denotes the potential, and
a dashed line denotes a correlation function, one from the bra and
another from the ket state. The vertices represent particles; each
vertex ranges over all particles in the nucleus.
If the correlation functions occupy
inequivalent positions, we get two distinct contributions from each diagram.
This leads to the total number of 16 contributions quoted in our previous work.
}
\label{fig:diagrams}
\end{figure}

In order to determine the ground-state energy
we need to compute the expectation value of both the kinetic
and the potential energy operators, as well as that of the unit matrix,
 to impose normalisation. The kinetic energy and norm matrices
are calculated straightforwardly, and we  concentrate therefore  on the
more complicated matrix element of the potential energy operator
\begin{equation}
\langle V \rangle = \langle\Psi_0\vert
\sum_{i<j} f^{\dagger}(ij) \sum_{k<l} V(kl) \sum_{m<n} f(mn)
\vert\Psi_0\rangle,
\end{equation}
where the indices in parentheses identify the particles on which the
corresponding operator acts. Many of the terms in the previous
expression are similar, and a convenient way of dealing with them is to make 
a diagrammatic analysis \cite{Bis91}
 and get the topologically distinct diagrams to expand the matrix elements.
The topologically distinct diagrams are 
 schematically indicated in Fig.~\ref{fig:diagrams}.

For each diagram we have to perform a summation over all occupied
single-particle states. Since we are dealing with saturated systems,
this sum is reduced to the sum over occupied harmonic oscillator
states times the respective trace of the spin and isospin operators,
where we must take account of all possible permutations of the labels
in the ket state, as explained in more detail in
\cite{Gua96}. Each permutation is split into a product of
transpositions with space, spin and isospin parts, $(ij)\equiv
(i^sj^s)(i^\sigma j^\sigma)(i^\tau j^\tau)$.  The spin and isospin
permutations can be represented by products of spin- and
isospin-exchange operators,
and are thus incorporated into the trace 
over spin and isospin. The final expression for matrix element related to a given 
diagram $d$ involving $N$ particles is
\begin{eqnarray}
\lefteqn{\langle\Psi_0\vert f^{\dagger}(ij) V(kl) f(mn)\vert\Psi_0\rangle_d}
\nonumber\\
&=&W_d \sum_{pqr}\sum^{A/4}_{i_1^s\dots i_N^s=1}\sum_P\epsilon_P
  \,\langle{\rm Tr}\ (\Theta^p_{ij}\Theta^q_{kl}\Theta^r_{mn}
   P^{\sigma}_{Pi_1^\sigma\dots Pi_N^\sigma}
    P^{\tau}_{Pi_1^\tau\dots Pi_N^\tau}) \nonumber\\
&&  \times \phi_{i_1^s}(1)\dots\phi_{i_N^s}(N)\vert
     {f^p}^{*}(r_{ij}) V^q(r_{kl}) f^r(r_{mn})
     \vert\phi_{Pi_1^s}(1)\dots\phi_{Pi_N^s}(N)\rangle.
\label{eq:Vdiag}
\end{eqnarray}
In this expression the $\phi$'s are the spatial parts of the
single-particle wave functions,
$W_d$ is the statistical weight of the diagram and 
$P^{\sigma}_{Pi_1^\sigma\dots Pi_N^\sigma}$ is the
representation of the spin part of the permutation
($P^\tau_{Pi_1^\tau\dots Pi_N^\tau}$ is the same for the isospin part).

The computation of the required traces is greatly simplified if
we work in the direct product representation since in this 
representation all the exchange and spin
matrices have the same structure, with only one non-null element
per row and column. In this manner the storage and matrix operations
can be performed very efficiently and, although the number of required
traces is large, the trace calculation may be done once
and stored for later use. It is worth
noting that the number of traces required in the V6 case is much
higher than in the V4 case due to the complicated structure of the
tensor operator. 

The last step in the calculation of the matrix elements is related
to the spatial integrals. There is a basic difference between
the V4 and V6 cases; in the former the operatorial structure does
not contribute to the spatial part of the calculation, but in
the latter a spatial contribution to the integrals arises.
This is due to the appearance of the spatial coordinates in the tensor 
operator, and we must therefore consider more types of
integrals. In order to perform the integrations we
expand the correlation functions in a set of Gaussians,
\begin{equation} 
f^p(r)=\sum_k c_k^p \exp (-\beta_k r^2) ,
\label{eq:Gaussex}
\end{equation}
with a factor $r^2$ for tensor components.
The Gaussian expansion has proven to give a very
 accurate representation of the
correlation function $f$ when both  negative and positive values
of $\beta$ \cite{Bis91} are included. 
In practice we use a set of around 10 Gaussians  with 
fixed length parameters
covering a sufficiently wide range of distances.
The integrals are computed (see \cite{Gua96})
by using a recurrence relation for all integrations apart from those involving
the coordinates in the potential; the remaining one-dimensional
integral, which involves
the potential, is then calculated numerically.
For the case of $^4$He we have also performed 
calculations by representing the functions $f^p$ on a grid in 
coordinate space, to check the accuracy of our results.

Since the expectation value of the Hamiltonian, as well as the norm,
are bilinear in the functions $f^p$, we can express all expectation
values as matrix elements, as obtained by substituting the expansion
(\ref{eq:Gaussex}) in expressions such as (\ref{eq:Vdiag}), and
isolating the contribution from a single pair of coefficients.  In this
way we can calculate the matrix elements of the Hamiltonian $
\mathcal{ H}_{mn}^{pq}$ and the norm $ \mathcal{ N}_{mn}^{pq}$. In
terms of these matrices one obtains for the expectation value of the
energy, depending on the set of coefficients $\{c_m^p\}$
(\ref{eq:Gaussex}), the formula
\begin{equation}
 E+E_{\mathrm{CM}}=\frac{{c_m^p}^* \mathcal{ H}_{mn}^{pq} c_n^q }
{{c_m^p}^* \mathcal{ N}_{mn}^{pq} c_n^q}, 
\end{equation}
where summation over repeated indices must be understood both in
numerator and denominator separately.

Optimising with respect to ${c_m^p}^*$ we obtain the
generalised eigenvalue problem
\begin{equation}
\label{eigenval}
\mathcal{ H}_{mn}^{pq} c_n^q 
=(E+E_{\mathrm{CM}}) \mathcal{ N}_{mn}^{pq} c_n^q,
\end{equation}
where the lowest eigenvalue is an approximation for the ground-state energy.
We must subtract the 
 centre-of-mass energy,
\begin{equation}
E_{\mathrm{CM}}=\frac{\hbar^2}{m}\frac{3}{4}\alpha^2,
\end{equation}
where $\alpha$ is the inverse length HO parameter, from this eigenvalue
to obtain the ground-state energy.
Given that the method is strictly variational, this energy may
also be optimised with respect to the harmonic oscillator parameter.

\subsection{Linear plus Jastrow correlations}
We shall also study a hybrid method which combines our TICI2 procedure
with its additive state-dependent correlations, and the Jastrow method
with its multiplicative state-independent correlations, thereby giving
rise effectively to a correlated basis functions (CBF) description.

It is known that Jastrow correlations are very well suited to treat
the effects of  short-range repulsion. 
On the other hand, the TICI2 description may well adapt itself
to represent the medium- and long-range correlations.
This motivated us to introduce
a Jastrow factor into the previous {\sl ansatz}, 
while maintaining the operatorial structure of the TICI2
correlation. Incorporating a scalar
Jastrow correlation between all pairs of nucleons in the TICI2 wave
function leads to the
following J-TICI2 wave function
\begin{equation}
\Psi_J({\bf r}_1,\ldots,{\bf r}_A) = \left( \prod_{i<j} g(r_{ij})\right)
\left(\sum_{k<l}\sum_p
f^p(r_{kl})\Theta^p_{kl}\right) \Phi({\bf r}_1,\ldots,{\bf r}_A). 
\label{jtici2}
\end{equation}
Due to limitations in computer power, we have restricted the
calculations to be presented below to a simplified Jastrow correlation
$g(r)$ defined in terms of a single
 Gaussian,
\begin{equation}
g(r)=1+a{\rm e}^{-br^2}.\label{eq:defg}
\end{equation}
 Our wave function now depends non-linearly on three
parameters, namely the harmonic oscillator parameter
 $\alpha$, and the width
$b$   and depth $a$ of the Jastrow function. In addition, it
 depends linearly on the functions $f^p(r)$ and $f^p(r)^*$. We
have to perform a variational search of the parameters to find the
minimum of the expectation value
$E_J \equiv \langle\Psi_J|H|\Psi_J\rangle/\langle\Psi_J|\Psi_J\rangle - E_{\rm CM}$.

\section{Results and discussion} \label{sec:results}

We shall investigate results obtained with the prescriptions
defined in the previous section for several V4 and V6
interactions, and for the nuclei $^4{\rm He}$,
$^8{\rm Be}$, $^{12}{\rm C}$ and $^{16}$O. 

We shall first look at the TICI2 method for tensor forces.
Afterwards, we discuss J-TICI2 results for the specific case of
$^4$He, both for V4 and V6 internucleon interactions. At the end of
this section we analyse the interplay of both kinds of correlations.
Preliminary versions of these results have been presented previously
in conference contributions \cite{CMT21,FB98}.

\subsection{TICI2 results for V6 interactions}

\begin{figure}
\begin{center}
\includegraphics[width=0.9\textwidth]{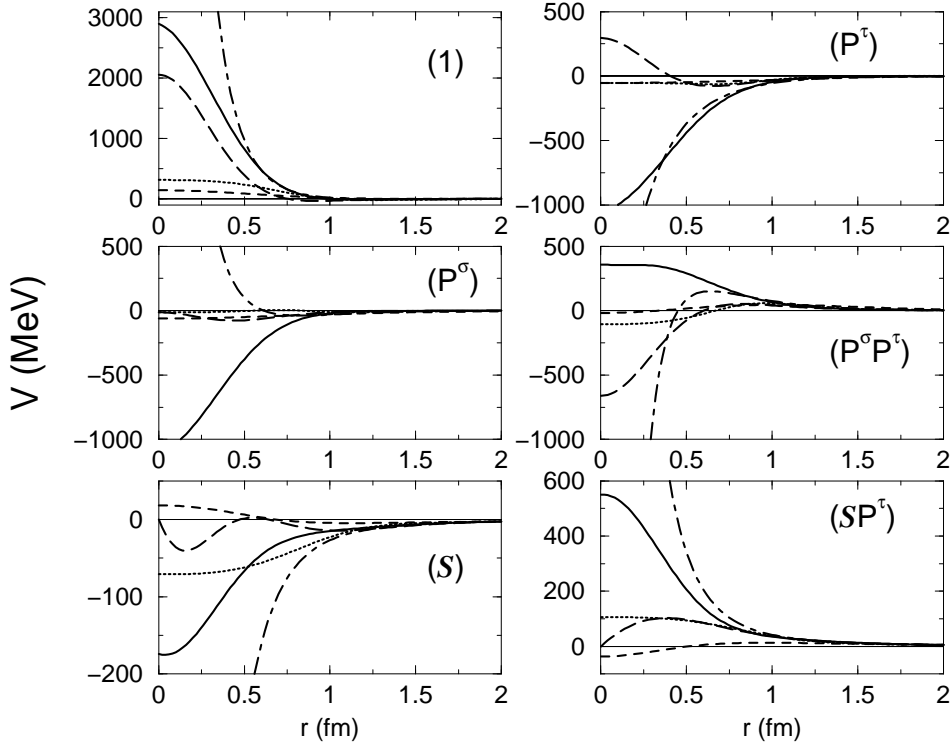}
\end{center}
\caption{A comparison between the V6 potentials used in this paper.
The horizontal axis in each panel shows $r$ (in fm), the vertical axis
the potential strength (in MeV). Each panel represents one operatorial
channel, for each of the operators used to expand the correlation functions,
as indicated in the figure. The solid line is AV14, the long-dashed
line is AV18. The dotted line is the SSC, and the medium dashed line
is the Gogny. The short-long dashed line is the Reid-V6
potential.  }
\label{fig:pot}
\end{figure}

We have performed calculations with a wide variety of interactions:
the Gogny potential \cite{GPDT}, the de
Tourreil-Sprung super-soft-core potential (SSC) \cite{SSC}, and the
Argonne potentials AV14 \cite{AV14} and AV18 \cite{AV18}.  In all the
cases we have used only the V6 part of the interactions, ignoring the
terms related to the spin-orbit and other different operators.  The
very different character of these potentials, specially at short
distances, can be seen in Fig.~\ref{fig:pot}. Here we include, in
addition to the potentials mentioned above, a modified version of the
Reid soft-core potential \cite{REID,Car88} (called Reid-V6), used
in later calculations for $^4$He only, and discussed further in
Sec.~\ref{subsec:JTICI2_V6}.

The introduction of tensor correlations in our scheme increases
dramatically the computer time needed in the calculations, as stated
in Sec.~\ref{sec:formalism}, and it has taken considerable effort to
deal with the case of $^{16}$O.  In Table \ref{tab:TICI2} we give the
results for the ground-state binding energy obtained for the
interactions mentioned above. We also list the value of the harmonic
oscillator parameter corresponding to the minimum.  Using our method,
we find that the V6 parts of the AV14 and AV18 potentials give a very
low value for the $^4$He binding energy. The nuclei $^8$Be, $^{12}$C
and $^{16}$O are even less bound, which is not what we expect to see.
The results obtained for the SSC and especially the Gogny potential
exhibit a more acceptable behaviour, showing at least a monotonic increase in
binding energy for increasing values of $A$.

\begin{table}
\caption{Binding energies (in MeV) of $0p$-shell nuclei using
TICI2 V6-like 
correlations, with the V6 part of the
 Gogny, SSC, AV14 and  AV18 interactions. The parameter $\alpha$ is the 
optimal harmonic oscillator 
parameter (in fm$^{-1}$).
} 
\label{tab:TICI2}
\begin{center}
\begin{tabular}{lllc}
\hline
Interaction & Nucleus  & $-E$& $\alpha$ \\
\hline
     & {$^4$}He    &  27.36& 0.70 \\
Gogny/V6 & {$^8$}Be    &  40.79& 0.58 \\
     & {$^{12}$}C  &  73.37& 0.61 \\
     & {$^{16}$}O & 128.68 & 0.64 \\
\hline
    & {$^4$}He   & 24.12  & 0.68\\
SSC/V6 & {$^8$}Be    & 25.98 & 0.54\\
    & {$^{12}$}C  & 39.64& 0.54 \\
    & {$^{16}$}O  & 63.55 & 0.55\\
\hline
     & {$^4$}He    &  14.77 & 0.59\\
AV14/V6 & {$^8$}Be   &  9.26  & 0.43 \\
     & {$^{12}$}C  &  10.50 & 0.41\\
     & {$^{16}$}O &  14.97 & 0.40 \\
\hline
     & {$^4$}He   &  15.40 & 0.61 \\
AV18/V6 & {$^8$}Be   &  11.13 & 0.47 \\
     & {$^{12}$}C &  14.96 & 0.46 \\
     & {$^{16}$}O &  23.76 & 0.46 \\
\hline
\end{tabular}
\end{center}
\end{table}

These results are very surprising in the light of the quality of our 
previous results
for V4 forces \cite{Gua96}, and merit further investigation. One 
extremely important difference is the competition between central and tensor
forces. In the V4 forces used in our previous work most of the binding
originates in the central channel. For 
all of the V6 forces used here,
most binding is obtained from the off-diagonal tensor force, 
i.e., the matrix element between the central and tensor-correlated channel. 
This matrix element competes with a strong repulsion in the central channel.
From Table \ref{tab:TICI2} and Fig.\ \ref{fig:pot} we see that the 
origin of
our problem seems to be related to the size of the 
repulsive core of the potential. We get
the most sensible result for the weak Gogny force, 
a slightly worse result for the more repulsive SSC potential,
and  very poor results for the AV14 and AV18 interactions.

\subsection{J-TICI2 results for V4 interactions}
\label{subsec:JTICI2_V4}

In order to shed some more light on these results we have investigated
the $^4$He system more deeply.  For this nucleus we can make
comparisons with other variational calculations and with Green
function Monte Carlo results.

We have performed calculations in the $^4$He system by using the
J-TICI2 wave function of Eq. (\ref{jtici2}).  With this {\em ansatz},
as already outlined in Sec.~\ref{sec:formalism}, we keep the
operatorial structure of TICI2 but add the flexibility of the Jastrow
correlations.

In order to investigate the origin of the different quality of TICI2
calculations for V4 and V6 interactions, we will start by considering
the simpler case of the V4 potentials that were studied in our previous
paper \cite{Gua96}, namely the Brink-Boeker B1 \cite{B1}, the
Afnan-Tang S3 \cite{S3} and its modified version MS3 \cite{MS3}, and
the Malfliet-Tjon MT~I/III and MT~V \cite{MT} potentials.  In
Table~\ref{tab:compv4}, we compare our J-TICI2 results for V4
interactions with the TICI2 ones \cite{Gua96}, and with those from a
recent Jastrow calculation \cite{Mariela} using SD correlations.  In
the case of the purely Wigner-like interactions, we also quote some diffusion
Monte Carlo (DMC) results \cite{Bis92b} which provide a benchmark for
the calculations.


\begin{table}
\caption{ Comparison between different results for the $^4$He binding
energy (in MeV) using several V4 potentials. S3$_{\rm W}$ is the
Wigner part of the S3 potential. Jastrow results refer to calculations
using state-dependent correlations containing two Gaussians.}
\label{tab:compv4}
\begin{center}
\begin{tabular}{lllll}
\hline
& TICI2       \protect\cite{Gua96} 
& J-TICI2 
& SD-Jastrow     \protect\cite{Mariela}
& DMC         \protect\cite{Bis92b}
\\
\hline
B1       & 37.86 & 38.28 & 38.27 \protect\cite{Bis92b} 
                                 & 38.32 $\pm$ 0.01 \\
S3       & 28.19 & 30.16 & 29.94 &                  \\
S3$_{\rm W}$
         & 25.41 & 27.20 & 27.21 & 27.35 $\pm$ 0.02 \\
MS3      & 27.99 & 29.97 & 29.70 &                  \\
MT V     & 29.45 & 31.21 & 30.88 & 31.32 $\pm$ 0.02 \\
MT I/III & 30.81 & 32.70 & 32.01 &                  \\
\hline
\end{tabular}
\end{center}
\end{table}

The main difference between the J-TICI2 and the SD-Jastrow 
calculations lies in the way  the state dependence of the correlations
is parametrised. In the case of SD-Jastrow, the
correlation factor is assumed to have the same operatorial
structure as the potential, and the product of correlated pairs
$\prod_{i<j}\left( \sum_p g^p(r_{ij})\Theta^p_{ij}\right)$ 
must be appropriately symmetrised to maintain
the Fermi statistics. In the SD-Jastrow results quoted in 
Table~\ref{tab:compv4}, each of the functions $g^p(r)$ was parametrised as
the sum 
of two Gaussians. 
On the other hand, in the J-TICI2 calculation the Jastrow
correlation factor is state-independent, so that all of the
operatorial dependence is contained in the pair correlation $f(r)$.
For computational reasons, the Jastrow factor in the latter case
was parametrised by a single Gaussian, as mentioned at
the end of the previous section.

In these model calculations we are trying to assess the effect of
superimposing the scalar Jastrow correlation on top of the linear pair
correlations.  We see from Table \ref{tab:compv4} that inclusion of
the scalar Jastrow factor lowers the ground-state energy (i.e.,
increases the binding energy) of the pure TICI2 results by around
1-2~MeV. The binding energies obtained for the Wigner-like potentials B1
and S3$_{\rm W}$ from the J-TICI2 method and from the use of
SD-Jastrow correlations are practically the same, but the former works
better for more realistic interactions with operatorial
structure. Especially interesting is the case of the MT potentials:
for the MT~V form, the J-TICI2 value is already better than the
SD-Jastrow one, despite the fact that the former includes only one
Gaussian whereas the latter includes two Gaussians in the correlation
functions $g$. This improvement can be explained by realizing that the
MT potentials contain a very short-range Yukawa core. The optimal
width of the Gaussian in the Jastrow correlation is very small ($b
\sim 5~{\rm fm}^{-2}$), which cannot provide at the same time a good
description for the long-range part of the correlation required to
modify the independent-particle reference state.  We also observe the
proximity of the J-TICI2 and DMC results, and we hence conclude that,
for these interactions, the J-TICI2 approach gives an excellent way to
treat the ground state of $^4$He.

In conclusion, the V4 results indicate the relevance of the hybrid
J-TICI2 description, assigning to each part of the wave function its
proper role: the accurate description of the short-range behaviour is
assigned to the scalar Jastrow factor, whereas the TICI2 part is
responsible for the medium-range correlations.

\subsection{J-TICI2 results for V6 interactions}
\label{subsec:JTICI2_V6}
This section deals with the most relevant application of the J-TICI2
{\em ansatz}, namely the consideration of realistic V6 forces.  In
addition to the potentials used before we have also considered a
modified version of the original channel-dependent Reid soft-core
(RSC) interaction \cite{REID}, which presents a much harder core than
the previous ones, and thus emphasises the role of short-range
correlations. Please note that we use the same channel-independent
Reid-V6 version of the RSC potential as has been used in the GFMC
calculations of Ref.\ \cite{Car88}.


\begin{table}
\caption{
Same as Table~\protect\ref{tab:compv4}, but referring to the  V6
parts of several realistic
potentials. The result attributed to  Ref.~\protect\cite{Car88} 
was obtained by adding the expectation values  of the
kinetic energy and the V6 part of the potential cited in that
paper, and hence it is not a fully optimised result. }
\label{tab:compV6} 
\begin{center}
\begin{tabular}{lccll}
\hline
&TICI2 \protect\cite{CMT21} &J-TICI2&
 VMC \protect\cite{CMT21} & GFMC\\
\hline
Gogny/V6 & 27.36 & 27.58 & 27.71 $\pm$ 0.06              & \\
SSC/V6  & 24.12 & 26.74 & 29.20 $\pm$ 0.12              & \\
AV14/V6 & 14.77 & 20.37 & 23.24 $\pm$ 0.08 
         & 24.79 $\pm$ 0.20 \cite{Car87}\\
AV18/V6 & 15.40 & 21.08 & 24.80 $\pm$ 0.09              & \\
Reid-V6 & 5.67  & 22.70 & 27.82 $\pm$ 0.12 
         & 28.30 $\pm$ 0.12 \cite{Car88}\\
\hline
\end{tabular}
\end{center}
\end{table}

Table \ref{tab:compV6} summarises our V6 results.  In addition to the
results obtained with the TICI2 and J-TICI2 descriptions, the table
also includes calculations with a V6 state-dependent Jastrow
correlation computed with the variational Monte Carlo (VMC) method, as
well as the V6 contribution to the energy computed with the Green
function Monte Carlo (GFMC) method for two interactions, AV14
\cite{Car87} and Reid-V6 \cite{Car88}, which provide the benchmark
results.  Our state-dependent Jastrow VMC calculations have been
carried out by using a separate Jastrow correlation function obtained
by solving Schr\"odinger-like equations for each operatorial channel,
as in Ref.~\cite{carlson_w}.

The pattern of V6 results is rather different from the V4 cases. The
effect of the inclusion of the state-independent Jastrow correlations on
top of the state-dependent TICI2 pair correlations is most impressive
for strongly repulsive potentials, such as the Reid-V6
\cite{REID}, in which the gain in binding energy is about
17~MeV. Clearly, with the exception of the extremely soft Gogny
potential, the J-TICI2 method has not converged to the level of
accuracy of the VMC (let alone GFMC) results. We note, however, that
one degree of freedom,
which has not yet been explored due to computational limitations, is
the use of a more general parametrisation for the Jastrow correlation
function $g$
in terms of multiple Gaussians. We expect some definite improvement in
energy, but it is as yet unclear whether this will provide us with the
bulk of the missing binding.  The nature of the binding (which, as we
shall see below, is mainly through off-diagonal components of the
tensor force) is such that we might expect that iterations of the
tensor correlations could also be important. The relevance of such
effects would only be addressed in a full TICC2 calculation.

\subsection{Analysis of the results}
In this section we present our thoughts about the origins of the
different behaviour of the TICI2 and the J-TICI2 results.  They are
based on the analysis of the overall size of the nucleus and on a
comparison of the contributions of the various operatorial parts to the
energy.

\begin{table}
\caption{TICI2 results ($E,\ \alpha$) and J-TICI2 results ($E_J,\
\alpha_J, \ a, \ b$) for $^4$He, using different interactions with V4
and V6 structures.   Binding energies are given in MeV, the HO parameters
$\alpha$ and $\alpha_J$ in fm$^{-1}$, the depth $a$ of the J-TICI2
Gaussian correlation function of Eq.~(\protect{\ref{eq:defg}}) is
dimensionless, and its width $b$ is measured in fm$^{-2}$.}
\label{tab:alpha}
\begin{center}
\begin{tabular}{l|cc|cccc}
\hline
& \multicolumn{2}{|c|}{TICI2} & \multicolumn{4}{c} {J-TICI2} \\
&$-E$&$\alpha$&  $-E_J$ & $\alpha_J$ & $a$ & $b$\\
\hline
\hline
{V4 Interactions}\\
B1      &37.86&0.73 & 38.28 & 0.77 & -0.41 & 1.8 \\
S3      & 28.19&0.72& 30.16 & 0.78 & -0.70 & 2.1 \\
S3$_{\rm W}$& 25.41&0.72
& 27.20 & 0.74 & -0.69 & 2.3 \\
MS3     &27.99 &0.71& 29.97 & 0.74 & -0.70 & 2.1 \\
MT V    &29.45&0.74 & 31.21 & 0.75 & -0.87 & 5.6 \\
MT I/III &30.81&0.74 & 32.70 & 0.75 & -0.88 & 5.0 \\
\hline
V6 Interactions\\
Gogny/V6    &27.36& 0.70& 27.58 & 0.72 & -0.29 & 1.9 \\
SSC/V6     &24.12& 0.68& 26.74 & 0.75 & -0.66 & 2.1 \\
AV14/V6    &14.77& 0.59& 20.37 & 0.71 & -0.93 & 2.7 \\
AV18/V6    &15.40& 0.61& 21.08 & 0.74 & -0.92 & 3.2 \\
Reid-V6    &5.67&0.43 & 22.70 & 0.72 & -1.05 & 2.4 \\
\hline
\end{tabular}
\end{center}
\end{table}

The first way of analysing the results is to compare the values of the
harmonic oscillator parameter as shown in Table \ref{tab:alpha}. We
see that, even for V4 interactions, it is the increase in $\alpha$,
that is provided by the inclusion of the Jastrow correlation factors
in Eq.~(\ref{jtici2}), which drives the improvements of the
energy. This need not surprise us; with a Jastrow correlation to kill
off the short-range repulsion, we can use the additive CI2
correlations to get the long-range physics right. Without Jastrow
correlations the CI2 correlation needs to play both roles, which works
better at smaller $\alpha$, where the nucleons are automatically
further apart.  The value of the HO parameter $\alpha_J$ for the
J-TICI2 {\em ansatz} does not change very much when considering
different interactions, ranging between 0.71 and 0.78~fm$^{-1}$ and
thus leads to similar overall sizes for the system. By contrast, the
values of the Jastrow parameters present a wider range of variation,
both in depth and in range, mainly due to the fact that the
short-range behaviour of the potentials is very different from one to
another.

\begin{figure}
\begin{center}
\includegraphics[width=0.6\textwidth]{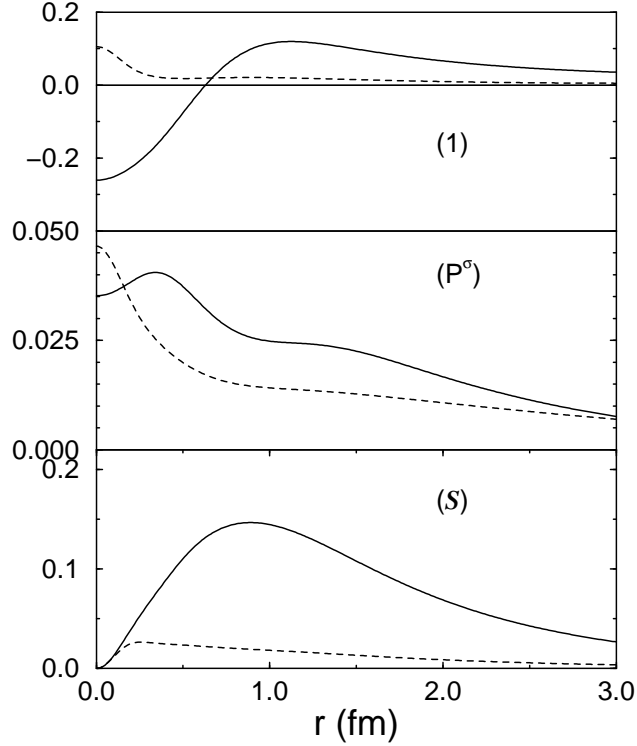}
\end{center}
\caption{A comparison between the TICI2 (solid line) and J-TICI2
(dashed line) correlation functions $f^p$, multiplied with $\exp(-\alpha^2
r^2/8)$, for $^4$He obtained from a solution to the generalised
eigenvalue problem (\ref{eigenval}). We have chosen the V6 part of the
AV14 potential as an example, and only plot the range for $r$ from 0
to 3 fm. The normalisation of each of the correlation functions is as
in the normalised ground-state eigenfunction, but we present separate
results for each of the three operatorial channels. }
\label{fig:V6}
\end{figure}

In Fig.~\ref{fig:V6} we compare the resulting correlation
functions $f^p$ for the AV14/V6 potential. Note that all these functions grow exponentially at large
radii, but these contributions are moderated by the harmonic
oscillator wave functions. For that reason we have multiplied these
correlation function with the factor $\exp(-\frac{1}{2A} \alpha^2 r^2)$, 
arising from the relative-coordinate part of the harmonic oscillator 
Slater determinant \cite{Gua96}.
 Sizes are related to the norm of the
related component in the ground-state eigenfunction. As expected we
see most change for the central correlation function. This has a node
for the case of the TICI2 calculation, due to the fact that it tries
to screen part of the hard-core contribution. For the  J-TICI2 method
this part of the correlation function is
very smooth and structureless, apart from a small bump at small
distances. We see that  the tensor
correlations are again rather smooth functions, which lends support to
our suggestion to include only state-independent Jastrow-factors in
our J-TICI2 method.

A slightly deeper understanding of the mechanism of binding can be
obtained by decomposing the Hamiltonian in channels, where in each
channel we use the combination of functions that was obtained from the
ground-state diagonalisation. This effectively reduces all matrices to
three-by-three ones (one index for each of the central , spin-dependent
and tensor-correlated channels) which allows for easier analysis. In a
more mathematical notation,
\begin{eqnarray}
H_{pq} &=& \langle\Phi'\vert \sum_{i<j}\Theta^p_{ij} {f^p}^* (r_{ij})
H \sum_{k<l}\Theta^q_{kl} f^q(r_{kl})\vert \Phi' \rangle
\equiv \sum_{m,n} {c^p_m}^* \mathcal{ H}^{pq}_{mn} {c^q_n}
,\nonumber\\
N_{pq} &=&
\langle\Phi'\vert \sum_{i<j}\Theta^p_{ij} {f^p}^* (r_{ij})
\sum_{k<l}\Theta^q_{kl} f^q(r_{kl})\vert \Phi' \rangle
\equiv \sum_{m,n} {c^p_m}^* \mathcal{ N}^{pq}_{mn} {c^q_n}
,
\end{eqnarray}
where $\Phi'$ is either the model state $\Phi$ itself or the
Jastrow-correlated model state, $\Phi'(\vec{r}_1,\ldots,\vec{r}_A)
=\prod_{i<j} g(r_{ij}) \Phi(\vec{r}_1,\ldots,\vec{r}_A)$, in the cases
of the TICI2 or J-TICI2 calculations, respectively.  The labels $p$
and $q$ run from one to three over central (1), spin ($P^\sigma$) and
tensor ($\mathcal{ S}$), respectively.

Since the norm matrix $N$ is diagonal, we can trivially take the square root,
perform the L\"owdin transformation, and obtain the symmetric  energy matrix,
\begin{equation}
\mathcal{E} = N^{-1/2} H N^{-1/2}.
\end{equation}
Let us compare these matrices in the case of the AV14 potential.
For the TICI2 wave function the energy matrix, subtracting $E_{CM}$
from the diagonal elements, has the value
\begin{equation}
\mathcal{E}_{\mathrm{TICI2}} = 
\left( \begin{array}{rrr}12.69&11.94&-96.38\\
11.94&212.54&-68.09\\
-96.38&-68.09&326.56 \end{array} \right).
\end{equation}
The eigenvector for the ground state is
$(0.960,0.033,0.278)$ with eigenvalue $-14.77\ \textrm{MeV}$.
Once we add the Jastrow correlation we find
\begin{equation}
\mathcal{E}_{\mathrm{J-TICI2}} = 
\left( \begin{array}{rrr}16.35&15.01&-109.66\\
15.01 & 294.05 & -74.89\\
-109.66&-74.89& 310.06 \end{array} \right).
\end{equation}
The eigenvector for the ground state is now
$(0.946,0.031,0.321)$ with eigenvalue $-20.37\ \textrm{MeV}$ .

In both cases the amplitude of the spin component is very small,
so that the influence of the spin channel is not of much relevance.
On the other hand, both diagonal matrix elements related to the central 
channel and to the tensor channel are {\em positive}, so that the
central-tensor coupling is the real source of negative energy.

\section{Conclusions}\label{sec:conclusions}

We have shown that the extension of our state-dependent TICI2
calculations to deal with tensor forces and tensor correlations is not
as straightforward as our previous work suggests. The reason seems to
lie mainly in the competition between the short-range repulsion
induced by the central force and the long-range correlations induced
by the tensor force.  The presence of unscreened potential
contributions, as shown in Fig.~\ref{fig:diagrams}, makes the TICI2
method not very well suited to deal with realistic interactions.  One
might well argue that a full CCM calculation, where only connected
diagrams appear, would alleviate this problem. Unfortunately, even
though some progress has been made \cite{cc2}, a coordinate-space
version of the translationally invariant CCM method has not yet been
completely and consistently formulated.

Even if we could perform such a calculation, we would still not
obtain full screening. Indeed, the method of Jastrow correlations
is much better suited for such purposes. The argument about the
source of repulsion (mainly of a central character) suggests
that the J-TICI2 method proposed here should be an excellent
starting point for describing light
nuclei with realistic interactions. Our results 
for $^4$He still give a few MeV
less binding than the much more elaborate VMC 
calculations and the (statistically) exact GFMC calculations. 

One
element that has not yet been fully explored is a better parametrisation
for the Jastrow function,
and definite conclusions about the quality of our approach
must be delayed until we are able to deal with a  more 
general central
Jastrow correlation. We are also actively investigating the best way to extend
the J-TICI2 calculations to heavier systems.

As a last comment, we note the conceptual similarity between our
approach and the correlated basis functions (CBF) description of
many-body systems, which may provide another avenue for future
development of hybrid methods.

\section*{Acknowledgements} 
This work is supported by the Committee of the European Community
under research contract No. ERBCHRXCT940456.  We acknowledge support
by DGICyT (Spain) under contract No. PB92-0820 (RG, JN and IM) and
contract No. PB95-0492 (AP).  RFB and NRW acknowledge support of a
research grant (GR/L22331) from the Engineering and Physical Sciences
Research Council (EPSRC) of Great Britain.

\end{document}